# High-frequency cyclicity in the Mediterranean Messinian evaporites: evidence for solar-lunar climate forcing

Vinicio Manzi[1,2*], Rocco Gennari[1,2], Stefano Lugli[3], Marco Roveri[1,2], Nicola Scafetta[4], B. Charlotte Schreiber[5]

[1] Dipartimento di Fisica e Scienze della Terra, Università di Parma, Parco Area delle Scienze, 157/a, 43124 Parma, Italy
[2] Alpine Laboratory of Palaeomagnetism (ALP), Via Madonna dei Boschi 76, 12016 Peveragno (CN), Italy.
[3] Dipartimento di Scienze della Terra, Universitá degli Studi di Modena e Reggio Emilia, Piazza S. Eufemia 19, 41100 Modena, Italy.
[4] Active Cavity Radiometer Irradiance Monitor (ACRIM) Lab, and Duke University, Durham 27708, NC , USA
[5] Department of Earth and Space Sciences, University of Washington, Seattle, WA 98195, USA
* Corresponding author's e-mail: vinicio.manzi@unipr.it

## ABSTRACT

**The deposition of varved sedimentary sequences is usually controlled by climate conditions. The study of two Late Miocene evaporite successions (one halite and the other gypsum) consisting of annual varves has been carried out to reconstruct the paleoclimatic and paleoenvironmental conditions existing during the acme of the Messinian salinity crisis, ~ 6 Ma, when thick evaporite deposits accumulated on the floor of the Mediterranean basin. Spectral analyses of these varved evaporitic successions reveal significant periodicity peaks at around 3-5, 9, 11-13, 20-27 and 50-100 yr. A comparison with modern precipitation data in the western Mediterranean shows that during the acme of the Messinian salinity crisis the climate was not in a permanent evaporitic stage, but in a dynamic situation where evaporite deposition was controlled by quasi-periodic climate oscillations with similarity to modern analogs including Quasi-Biennial Oscillation, El Niño Southern Oscillation, and decadal to secular lunar- and solar-induced cycles. Particularly we found a significant quasi-decadal oscillation with a prominent 9-year peak that is commonly found also in modern temperature records and is present in the contemporary Atlantic Multidecadal Oscillation (AMO) index and Pacific Decadal Oscillation (PDO) index. These cyclicities are common to both ancient and modern climate records because they can be associated with solar and solar-lunar tidal cycles. During the Messinian the Mediterranean basin as well as the global ocean were characterized by different configurations than at present, in terms of continent distribution, ocean size, geography, hydrological connections, and ice-sheet volumes. The recognition of modern-style climate oscillations during the Messinian suggests that, although local geographic factors acted as pre-conditioning factors turning the Mediterranean Sea into a giant brine pool, external climate forcings, regulated by solar-lunar cycles and largely independent from local geographic factors, modulated the deposition of the evaporites.**

## 1. INTRODUCTION

The Messinian salinity crisis (MSC) is a complex geological event that resulted in the transformation of the Mediterranean Sea into a giant brine pool where a dramatic hydrological and biological crisis occurred at about 6 Ma (Hsü et al. 1973). Its geological record (Fig. 1) consists of sulfate and halite deposits with a volume of more than 1 x $10^6$ $km^3$ (Hsü et al. 1978; Montadert et al. 1978; Rouchy and Caruso, 2006) accumulated in less than 700 ka (Krijgsman et al. 1999) in both shallow-water and deep-water settings (Roveri et al. 2008; CIESM, 2008). The climatic conditions and forcing mechanisms that were active at different temporal scales during the evaporite deposition are still largely unknown. A consensus has been reached in considering that evaporite deposition was controlled by the variation of Earth's orbital parameters and particularly by precession that drove an ~ 20 kyr alternation of drier vs. more humid conditions (Fig. 2; Krijgsman et al. 1999; van der Laan et al. 2006; Gargani et al. 2008; Roveri et al. 2008; CIESM, 2008; Manzi et al. 2009; Lugli et al. 2010). These mechanisms explain the decameter-scale lithological cyclicity given by the rhythmic alternation of evaporitic and fine-grained terrigenous bodies. Much more uncertain is the origin of centimeter to millimeter thick evaporite rhythmites observed in the evaporites of all MSC stages; they represent the only cyclical pattern recognizable in the halite deposits formed during the very short interval (less than 50 kyr) of the MSC acme (6.0-5.5 Ma, stage 2; CIESM, 2008; Fig. 2).

Here we present a detailed study of the small-scale lithological cyclicity observed in two evaporite sections from the MSC stage 2 (CIESM, 2008) recorded in the Caltanissetta basin of Sicily (Figs. 3, 4), which represent a unique archive for assessing the paleoenvironmental and paleoclimatic conditions controlling and modulating the deposition of large salt bodies in the deep Mediterranean basin. On the other hand, this study also provides important insights on the characteristics of Earth's climate system during the Late Miocene, i.e. prior to the Northern Hemisphere glaciation (Raymo et al. 1989; Roveri and Taviani, 2003).

## 2. THE STRATIGRAPHY OF THE MESSINIAN SALINITY CRISIS

After decades of controversy (see Rouchy and Caruso, 2006 for discussion) an international agreement has been reached on the stratigraphy and time sequence of the Messinian salinity crisis in the Mediterranean (CIESM, 2008). The discussion of alternative interpretations supported by other research teams is beyond the aim of this paper; here we simply adopt the stratigraphic framework proposed by CIESM (2008) and its most recent refinements (Roveri et al. 2008, Lugli et al. 2010; Manzi et al. 2009; Manzi et al. 2011) that subdivide the Messinian salinity crisis into three main evolutionary stages, each one characterized by peculiar evaporite deposits recording different hydrological and paleoceanographic conditions.

The ***first stage*** (5.96-5.6 Ma; Figs. 2, 3), dominated by sulfate evaporites, corresponds to the deposition of massive bottom-grown selenite (Primary Lower Gypsum; PLG; Roveri et al. 2008) in semi-isolated, marginal sub-basins, whereas only organic-rich shale and dolomitic limestone accumulated in deeper and/or more open settings (Manzi et al. 2007). The PLG evaporites were deposited in small and moderately deep (< 200 m), episodically oxygenated basins, and precipitated from a relatively homogeneous Atlantic-fed water body with a restricted outflow and a significant contribution of continental waters (Lugli et al. 2007; Lugli et al. 2010).

The MSC acme occurred during the ***second stage*** (5.6-5.55 Ma; stage 2.1 of CIESM, 2008; Figs. 2, 3) dominated by $CaCO_3$-NaCl-K salts, and characterized by a strong phase of erosion and tectonic activity. Its sedimentary record consists of both clastic deposits, derived from the erosion and resedimentation of stage 1 PLG evaporites (RLG; Resedimented Lower Gypsum; Roveri et al. 2008), and of primary evaporites; the latter consisting of thick and extensive primary halite and gypsum cumulate bodies that were mainly deposited in the deeper settings. This stage comprises the TG14 and TG12 glacials and was triggered by a combination of pan-Mediterranean tectonic and climatic factors that caused a significant reduction of the Atlantic connections and the likely blockage of the Mediterranean outflow.
The ***third stage*** (5.55-5.33 Ma; stage 2.2 of CIESM, 2008; Figs. 2, 3) is characterized by the deposition of $CaSO_4$ evaporites in both shallow and deep settings alternating with hypohaline sediments, corresponding respectively to the Upper Gypsum (UG) and Lago Mare (LM) deposits. Precipitation of UG selenite occurred from a large, residual water body only partially connected with the Atlantic, as suggested by the depleted strontium isotope ratio with respect to the coeval oceanic waters (Flecker and Ellam, 2006; Manzi et al. 2009). This stage was followed by the return of fully marine conditions in the Mediterranean (Hsü et al. 1978; Van Couvering et al. 2000).

## 3. THE RECORD OF THE SALINITY-CRISIS ACME (STAGE 2) IN THE CALTANISSETTA BASIN

The Messinian primary evaporites are characterized by a pervasive, small-scale lithologic cyclicity. In the selenite bodies deposited during MSC stages 1 and 3 such a small-scale cyclicity is marked by the centimeter-thick rhythmic alternation of gypsum selenite crusts and shale or limestone veneers (Fig. 3; Lugli et al. 2010; Manzi et al. 2011;). Small-scale cyclicity is even more evident in stage 2 primary evaporitic deposits (mainly halite and gypsum cumulates) consisting of thin (centimeter) to very thin (millimeter) evaporite-shale alternations.
Here we focus on the halite and gypsum cumulate deposits of the Caltanissetta basin (Sicily), which likely represent the most continuous onshore records of the crisis acme.
We measured 120 lithological cycles for an overall thickness of 18 m (Figs. 4, 5) in the halite deposits of the ***Realmonte salt mine*** (37°17'47.31" N; 13°28'23.20" E; Lugli, 1999). Each cycle is tripartite and consists of a basal millimeter-thick shale veneer, a thin (millimeter to centimeter) anhydrite ($CaSO_4$) and/or polyhalite ($2CaSO_4 \cdot MgSO_4 \cdot K_2SO_4 \cdot H_2O$) layer, capped by a decimeter-thick massive halite (NaCl) bed. The thickness of each cycle is somewhat variable, with an average value of 15.1 cm (standard deviation 4.7 mm obtained on 120 cycles), a maximum of 30 cm, and a minimum of 1 cm, but 80% of the cycles show a thickness ranging from 10 to 20 cm. The massive halite consists of primary halite cumulates (i.e., crystals forming at the pycnoclinal interface and then settling to the bottom) formed by skeletal hoppers showing further vertical overgrowth (chevron) that occurred at the bottom of the basin after initial growth at the brine surface (Lugli, 1999).
A second section was measured at ***Pasquasia*** (37°31'57.60" N, 14°12'10.61" E; Selli, 1960) in a laminar gypsum ($CaSO_4 \cdot 2H_2O$) unit (Figs. 4, 6) made up of 1179 primary gypsum laminae and subordinately by fine-grained clastic gypsum. According to Selli (1960), these deposits pass laterally to a halite body, which is coeval with that of the Realmonte mine (Fig. 4). The gypsum laminite unit is a millimeter-scale rhythmic succession of gypsum laminae (Ogniben, 1957). The thickness of individual laminae varies between a minimum of 0.4 mm and a maximum of 6.8 mm, with an average value of 2.6 mm (standard deviation 0.9 mm obtained for 979 laminae). The gypsum laminae consist of inversely graded (i.e. with crystal size increasing toward the top of each lamina; Fig. 6c) cumulates of fine-grained granular crystals separated by very thin veneers of organic matter (less than 1 mm). These features suggest that each lamina represents a discrete basin-scale evaporitic event, very similar to the laminar gypsum cumulates occasionally present in modern salt works.
The cumulitic laminae are arranged to form up to 54 lamina sets (i.e., group of laminae; numbered from the bottom of the section in Fig. 6) formed by a highly variable number of laminae (3 to 65) separated by centimeter-thick brown shale horizons; less-prominent millimeter-thick shale veneers allow a further subdivision for a total of 79 lamina subsets (internal subdivision are indicated in Fig. 6, when present). The lamina sets from 1 to 51 are in turn arranged in 14 clusters (i.e., groups of lamina sets; numbered from the bottom of the section in Fig. 6), separated by decimeter-thick fine-grained (from very-fine sand to silt) clastic gypsum beds (named from the bottom in alphabetical order in Fig. 6) likely deposited by subaqueous low-density gravity flows. The base of the clastic beds is commonly sharp, but erosional or dissolution features have not been observed in the underlying cumulate laminae.
Three uppermost lamina sets (named 52, 53, and 54) are locally altered by modern pedogenesis, thus hampering the accurate counting of laminae. Nevertheless, 500 to 600 additional laminae can be roughly estimated for this interval. Thus, the whole section is estimated to contain 1700 to 1800 laminae. Individual lamina sets thicken upwards due to a progressive decrease in frequency and thickness of the clastic beds, suggesting greater stability of evaporitic conditions through time.

## 4. TEMPORAL HYPOTHESIS: THE PALEOENVIRONMENTAL MEANING OF SMALL-SCALE EVAPORITE CYCLICITY

Following the principle of geological uniformitarianism (Hutton, 1785; Hutton, 1795; Lyell, 1830), the basic assumption of this study considers that, similarly to what commonly occurs in modern evaporitic environments, each discrete elementary evaporitic event recognized in the studied sections of Realmonte (shale-anhydrite-halite triplets) and Pasquasia (shale-gypsum couplets) records annual dry-humid environmental variations. This basic assumption is also substantiated by further considerations.
The deposition of the evaporites during MSC stages 1 (PLG) and 3 (UG) was mainly controlled by precession producing ~ 20 kyr evaporite-shale alternation, respectively related to dry vs. humid conditions (Fig. 2; Krijgsman et al. 1999; van der Laan et al. 2006; Roveri et al. 2008; CIESM, 2008; Manzi et al. 2009; Lugli et al. 2010). In contrast, the primary evaporite unit of stage 2, the deposition of which has been constrained to fall within a 50 kyr time window (Fig. 3), does not show any clear internal large-scale subdivisions or interbedded shale horizons (Fig. 4) that may allow the identification of precession-related cyclicity. All of the largest halite bodies of the Caltanissetta basin (fig. 4) are characterized by a similar threefold internal organization given by an intermediate unit enriched in more soluble salts, mostly kainite ($4MgSO_4 \cdot 4KCl \cdot 11H_2O$) and minor amounts of carnallite ($MgCl_2 \cdot KCl \cdot 6H_2O$) and bishofite ($MgCl_2 \cdot 6H_2O$), sandwiched between two halite units (Decima and Wezel, 1971; Lugli et al. 1999). This quasi-symmetrical cycle of halite passing to K-Mg salts and returning to halite could be related to a large-scale evaporation cycle that reached its acme with the deposition of the more soluble salts. This trend is supported by the facies distribution showing a progressive shallowing in the depositional facies, up to local ephemeral desiccation conditions in the lower portion (Lugli et al. 1999). This desiccation was followed by a less pronounced deepening trend based on the increase of accommodation for the upper unit (Lugli et al. 1999; Roveri et al. 2008). It follows that the whole salt unit could have accumulated within a single precessional cycle, or even more likely, by deposition during its dry hemicycle, thus spanning a maximum interval of 10 kyr. Similarly, the gypsum cumulate unit (Manzi et al. 2009) lacks a clear internal subdivision and is characterized in its intermediate to upper portion by the maximum development of the cumulate facies.
The overall number of small-scale cycles counted in the gypsum cumulates of Pasquasia section is between 1700 and 1800. In the Realmonte salt mine, the maximum thickness of halite can be estimated at around 400 m (Fig. 4); thus, extrapolating the data of our section, a rough total number of 2500 cycles can be calculated for the whole salt unit. Assuming a duration for each small-scale cycle of 10 yr, the cumulate unit could exceed the duration of one entire precessional cycle; thus a shorter duration probably should be envisaged.
The comparison of the Sicilian evaporites with modern (e.g. Lake Mac Leod, Western Australia; Warren, 2006) and ancient analogues (Anderson, 1982; Anderson and Dean, 1995; Leslie et al. 1996; Kirkland et al. 2000; Kirkland, 2003) suggests that the small-scale cyclicity observed in Messinian evaporites may record annual, pluriannual, and decadal high-frequency alternations of dry and humid phases.
Nevertheless, both the Realmonte triplets and the Pasquasia couplets clearly describe discrete evaporitic events starting from undersatured conditions and ending only after strong evaporative conditions, without any evident interruptions. In fact, the gypsum laminae of Pasquasia are characterized by a peculiar inverse gradation of cumulitic crystals and do not show any dissolution or shale deposition that could suggest dilution events in the basin. In the same way, the shale-anhydrite-halite triplets of the Realmonte salt mine describe events of progressive saturation of the water mass, without any dilution events. Deposition of halite and gypsum cumulates can be related to evaporitic environments characterized by seasonal alternation of dry and humid environmental conditions, as occurring in a single year in modern environments. Consequently, we suggest that the evaporite-shale couplets of stage 2 can be interpreted as varves, and we assume as a working hypothesis for the statistical analyses that they have annual duration. Evaporite precipitation occurred during the dry season, whereas during the humid season argillaceous sediments carried by runoff freshwater accumulated at the bottom of the basin.

Moreover, the sedimentation rate of 15 cm/yr obtained for the halite cumulates cycles of the Realmonte salt mine has a magnitude comparable to the one calculated for the deposition of halite chevrons in modern salinas (10 cm/yr; Schreiber and Hsü, 1980). Depositional rates of gypsum cumulates are more difficult to assess, inasmuch as no modern analogues are available; nevertheless, we observed minor gypsum cumulites forming sporadically in modern, human-operated salinas in Spain (Cabo de Gata) and Sardinia (Cagliari), suggesting annual depositional rates varying from a few millimeter to a few centimeter, strongly depending on the local hydrological conditions. Simple counting of evaporite varve enabled us to obtain an age model for these deposits (Muñoz et al. 2002); no additional calculations were performed to convert thickness into time or to calculate the sedimentation rate. Erosional and/or other features indicating significant hiatuses have not been observed in these sections; this suggests that the deposition took place in a very low-energy environment able to provide a continuous sedimentary record.  Unfortunately, the Realmonte and Pasquasia sections cannot be correlated bed by bed because they consist of different lithologies (halite-anhydrite-shale triplets vs. gypsum-shale couplets and represent different time spans (120 vs. 1700-1800 cycles). Although we cannot provide the spectacular correlation shown for the sulfate varves of the Permian Castile Formation (Anderson, 1982) here, the internal organization of laminae formed by different lithologies demonstrates that they have the same meaning as annual varves.

## 5. MATERIALS AND METHODS: THE STUDY OF EVAPORITE VARVES

In modern evaporative basins, the salinity and the density of the water mass, and the consequent evolution of the brines are strictly controlled by the effective evaporation rate (Warren, 2006), which, in turn, depends on many environmental factors such as wind velocity, air temperature, water temperature, and solar irradiance. Thus, the volume of the evaporite seasonally accumulated in a basin can be considered by far one of the best proxies for the short-term climate variability. In order to reconstruct the shorter-term paleoclimatic oscillation during the acme of the Messinian salinity crisis, we analyzed the thickness variation of evaporite varves and the return time of the dilution events which are represented by thicker shale horizons and/or low-density gravity flows.

We performed statistical analyses using power-spectrum estimations by the maximum entropy (all poles) method (MEM; see Courtillot et al. 1977 and Press et al. 1997 for further information) and the more traditional Lomb Periodogram (LP). MEM is complementary and has some advances over traditional power-spectrum periodogram methods such as LP and fast Fourier transforms, because it is based on an autoregressive algorithm that can give an accurate representation for an underlying power-spectrum function by looking for mathematical poles producing thin delta-like functions at the statistically relevant frequencies. Other techniques produce wide bell-shaped peaks whose highs give the power associated with the amplitude of a given cycle. The MEM delta-like peaks may allow robust separations of close frequencies, and the technique may work well for short sequences as well. The maximum number of frequencies that may be detected is called “order of poles” and cannot be larger than half-length of the record. When this technique is used with poles M ranging from N/5 to N/2 (where N is the length of the data), the methodology is optimized in detecting possible cyclicities at the low-frequency range such as decadal and multi-decadal scales (Courtillot et al. 1977). In the following we use the above two alternative spectral-analysis methodologies for reciprocal validation and use M = N/2.

## 6. RESULTS OF SPECTRAL ANALYSIS

Spectral analysis of the cyclic halite at Realmonte mine (Fig. 7A) contains 120 values that, according to our hypothesis, correspond to 120 consecutive years. The power spectra are shown in Figure 7B. The most prominent peaks, with respect to red-noise confidence levels relative to the Lomb-Periodogram estimates (90% and 95%; Fig. 7B), are those with periods of 3.2 years and 9 years. Other possible peaks are observed at 7.6, 11.5, 15, and 25-27 years, but their significance and stability is more uncertain. This record is too short to accurately determine the significance of the longest detected periodicities or of the smaller peaks.
The record of thickness variation for the gypsum laminae in the Pasquasia section (Fig. 7C) contains 1179 values that, according to our working hypothesis, correspond to 1179 consecutive years. The record appears to be segmented into five groups because some intervals cannot be resolved due to alteration; more precisely, for these intervals it was possible to count only the number of laminae, but not their individual thicknesses, due to the lack of fresh cuts perpendicular to the lamination. For these reasons only the average thickness was considered for the altered intervals. The first uninterrupted interval is made up of 444 data points, and it is the longest uninterrupted record available. The power spectrum of this first part of the sections is shown in Fig. 7D. Both spectra show numerous peaks against the estimated red-noise confidence level having dominant peaks in the ranges of 3.5-5.5, 6, 7.3, 8.6, 11.2, 13, 17.6, 19.6, 51, and 83 years. However, the most prominent variability is observed at about 4 years and at 20 years, plus a significant variability within the 50-100 years scale.
Finally, we investigated the possible occurrence of regular periodicities in the recurrence of dilution events (floods, rainfalls) in the evaporitic basin through the probability distribution analysis using Gaussian kerner density estimates performed on the duration of each of the 79 lamina subsets (see internal subdivision of laminasets 1 to 51 in Fig. 8) obtained by counting the number of cumulate layers separating two successive shale horizons (Fig. 7E). The analysis shows significant probability peaks at periods of 4, 7, 9, 11, 15, 20, 24.5, and 32-35 yr, with the most prominent peaks observed again in the 7-11 years range (Fig. 7f).
Furthermore, an evolutive spectral analysis was performed on the thickness variation for both of the Realmonte and Pasquasia cumulate series: the Morlet wavelet spectra (Fig. 8). The results show the typical interference-beat patterns among the different frequencies obtained from spectral analysis. In the Realmonte section (Fig. 8A) an approximate 60-year beat cycle can be recognized around the 8-12 years period range, while in the Pasquasia section the ENSO like variability around 3-7 year period range is more clearly highlighted (Fig. 8b).

## 7. DISCUSSION

### *7.1 Short-Term Modern Climate Forcing*

Earth’s climate is a very complex geophysical system controlled by the interaction of several terrestrial and astronomic forcing, producing an intense cyclical variability at different time scales that can be reduced or amplified according to the physiography of each basin.
In the study of climate-sensitive environments that can record the shorter-term (annual and pluriannual) climate variabilities, like lacustrine or marine evaporitic environments, the most important climatic variations can be considered in terms of air and water temperature, precipitation rate, and intensity and recurrence of precipitations events (Warren et al. 2006 and references therein).
In modern times several climate variations are recognized based on particular periodicities (Table 1).
The Quasi-Biennial Oscillation (QBO) has a mean period of about 2.3-2.5 yr (Baldwin et al.

2001); periodicities between 3 and 7 yr are commonly found in the El Niño Southern Oscillation (ENSO; Trenberth, 1997); an oscillation at about 9 yr is clearly present in the global surface temperature records since 1850 and can be related to a lunisolar long-term cycle (SLC; Scafetta, 2010, 2012a, 2012b, see pages 35 and 36 in the Supplement); periods at about 10-13 yr are usually related to the 11 yr sunspot-number solar cycle (SSC; Hoyt and Schatten, 1997); 19 and 25 yr periods are commonly associated with the 22 yr Hale solar magnetic cycle (HSC; Hoyt and Schatten, 1997) and to the 18.03 yr Saros cycle, the 18.6 yr luni-solar nodal cycle, and/or the 18.6 nutation cycle, which may influence the Arctic climate and likely regulate a bidecadal North Atlantic Oscillation; (NAO; Yndestad, 2006).
Clear quasi 20-yr cycles in the global surface temperature in synchrony with astronomical cycles also have been observed (Scafetta, 2010, 2012a).
Longer-term oscillations with periods of 50-100 yr are also found in numerous climate records (Wyatt et al. 2011) and are commonly associated with the Gleissberg solar cycle (GSC; Hoyt and Schatten, 1997). For example, 60-64 yr and 80-90 yr cycles have been found in solar-related records over the past 10 ka (Ogurtsov et al. 2002) from different proxies such as $^{10}$Be from polar ice (1010-110 B.P.) and $^{14}$C data from tree rings (9758-65 B.P.).
Cycles with quasi 50-70-year periods are found in several climatic records (Wyatt et al. 2011), particularly in the global surface temperature since 1850 (Scafetta, 2010, 2012a) and are clearly present in the North Atlantic Oscillation indexes since 1700 (Mazzarella and Scafetta, 2012).
Spectral peaks that fall at 12.7, 26.5, 34.4, and 57.3 yr periodicity for temperature and a marked 90-yr maximum for precipitation have been found to characterize the western Mediterranean climate in the last four centuries (Camuffo et al. 2010).

### *7.2 Modern vs. MSC Climate-regulation Mechanisms*

As commonly observed in modern records (see Scafetta 2010, 2012a and references therein), power spectra obtained for MSC deposits (Fig. 7B, D, F) document well-developed interannual climate cyclicity. The frequency peaks obtained for the Messinian acme salts reveal periodicities not exactly equal to each other and may be only approximately accurate. However, we observed a rough similarity with the typical major frequency ranges reported in modern climate records; because both the Realmonte and Pasquasia sections represent marine Mediterranean records, the observed climatic oscillations can be representative of comparable temperature and/or precipitation cycles.
Our data suggest that both the thickness of the evaporite varves (fig. 7B, D) and the occurrence of fluvial flood-related deposits (Fig. 7F) vary cyclically with periodicities significantly corresponding to those reconstructed for the last 500 yr for modern temperature and precipitation in the western Mediterranean (Camuffo et al. 2010) that are currently related to both astronomical (lunar and solar cycles) and oceanic cyclicities (QBO, ENSO).
The influence of the El Niño/Southern Oscillation (ENSO) has been suggested as a main forcing mechanism for Messinian evaporitic deposits (Galeotti et al. 2010), thus implying ENSO variability during the Late Miocene and its possible teleconnections. Our work, based on a much longer time series (1200 yr vs. 250 yr of Galeotti et al. 2010) confirms previous findings. Moreover, it also documents that climate forcing other than the ENSO-like oscillations and not previously considered were active in the late Miocene and played an important role in controlling the Messinian evaporite systems.
The most prominent finding of our spectral analyses (figs. 7B, F) is that, in addition to the QBO and ENSO frequency band (2-5 years), a strong periodicity at the decadal scale around about 9 years can be detected Figures 7b and 7f). This decadal cycle, which is quite evident in several temperature records, can be related to major solar oscillations and lunisolar tidal cycles, which can modulate ocean currents and the albedo through a direct forcing of the cloud system (Scafetta, 2010, 2012a, 2012b Appendix, 35-36). Thus, numerous astronomical oscillations would influence climate, as is already known to happen for tidal dynamics (Ehret, 2008; Wang et al. 2012). Variations in the ocean circulation are well known to drive and influence the global climate and would influence Europe and the Mediterranean region as well. A significant common 9 yr cyclicity in the power spectrum of the modern Atlantic Multidecadal Oscillation (AMO) index (155 yr; 1856-2011) and the Pacific Decadal Oscillation (PDO) index (111 yr; 1900-2011) is shown in Figure 9. Other significant spectral peaks at 2-3 yr and 4-8 yr, and a possible multidecadal peak at about 60 yr, approximately agree with the spectral ranges depicted in Fig. 7.
As for the longer-term findings, the 85-90 yr and 60-110 yr oscillations found in both the Realmonte and the Pasquasia records also matches reasonably well with the 90 yr rain periodicity in the Western Mediterranean climate of the last four centuries (Camuffo et al. 2010). Note that the Realmonte record is quite short (120 yr), and the existence of a quasi-secular cycle can be inferred only from the fact that the analysis highlights a multi-decadal, quasi-secular modulation that could be consistent with an oscillating pattern, which is more clearly measurable in the longer record of Pasquasia.

### *7.3 Main Implications for the Messinian Salinity Crisis*

During the late Miocene the paleoenvironment of the Mediterranean underwent an important geodynamic event implying the tectonic reorganization of the Eurasia-Africa plate boundary and significant physiographic changes (Duggen et al. 2003; Jolivet et al. 2006; Govers et al. 2008). However, the Messinian paleogeography and paleoceanography of the Mediterranean area remains under debate (Blanc, 2000; Meijer and Krijgsman, 2005, Govers et al. 2008; Gargani and Rigollet, 2008), especially concerning basin depth, sea level, and land distribution (see Fig. 1B). One certain paleocenographic consequence was the reduction of the connections with the Atlantic Ocean, as witnessed by the decrease of the Sr isotope ratio (Flecker et al. 2002) and the progressive development of bottom-water anoxia. Nevertheless, a continuous oceanic influx into the Mediterranean is required for the large volume of ions ($Ca^{2+}$, $SO_4^{2-}$, $Na^+$, $Cl^-$) needed to precipitate the evaporites during the three crisis stages (Manzi et al. 2009).
The primary evaporite record of the Realmonte and Pasquasia sections suggest that the local MSC acme (stage 2) occurred within a single precessionally driven glacial interval (CIESM, 2008; Manzi et al. 2009) and had an overall duration of a few thousand years. This very short time window likely corresponds to the peak of a glacial episode (isotope stages TG12 or TG14; Roveri et al. 2008); the associated lowering of global sea level would have further reduced the connections between Atlantic Ocean and the Mediterranean Sea, thus favoring the progressive increase of water saturation up to a basin-wide halite phase. This situation could also have increased the sensitivity of the Mediterranean to paleoceanographic changes, thus amplifying the effects of annual and pluri-annual climate oscillations.
Our results indicate that the Mediterranean climate during the MSC peak was characterized by climatic oscillations similar to those observed during the past few centuries, likely regulated by astronomical cycles related to the Moon and to the natural oscillations of the Sun and the heliosphere (Scafetta, 2010; 2012a, 2012b, 2012c, 2012d, Wang et al. 2012). Evaporite deposition was caused by precipitation at the pycnocline interface induced by intense evaporation during the driest season and hampered by water-mass dilution related to precipitation and fluvial runoff increase during the wet season (Fig. 10).
Consequently, if we consider the Mediterranean in a simplistic way, we can argue that solar irradiation (plus wind intensity) in the dry season vs. precipitation rate in the wet season would have been the main factors controlling evaporite accumulation. As for the annual scale, the alternation of wetter vs. drier climate conditions may also have influenced the deposition of evaporites in the Mediterranean basin at a pluri-annual to decadal scale. In our opinion, during dry phases the strong evaporation rate induced a more severe negative hydrological budget favoring a strong input of Atlantic Ocean water, whereas during wet periods the increase of precipitation and continental runoff likely triggered a significant reduction of the ocean exchange. This would have produced stratification of the water column and anoxia at the basin floor.

### *7.4 Implications for the Origin of Modern and Ancient Climate Forcing*

A more general implication of our study concerns the effective role of ENSO during the past, which is one of the most debated paleoclimatic issues. The existence of "permanent" El Niño-like conditions during warm periods has been postulated with analogies with the Pliocene Warm Period (Wara et al. 2005; Fedorov et al. 2006). However, this hypothesis has been questioned because ENSO-like oscillations have been reconstructed from the $\delta^{18}O$ anomalies of a coeval coral record that are considered to be more appropriate high-resolution proxies of sea-surface temperature (Watanabe et al. 2011). Moreover the extension of "intermittent" ENSO conditions back in the geological record, as ancient as the Late Cretaceous, has been suggested by several authors (Davies et al. 2012 and reference therein).

This interpretation is also in agreement with our observations for the Miocene, but our results further suggest new insights about the existence and the physical origin of longer natural climatic oscillations.

It is noteworthy that during ancient periods characterized by varied continental distribution, size and geometry of the oceans, hydrological connections and ice-sheet distribution, climatic periodicities were similar to those observed today. The most plausible explanation for these findings is that climate oscillations at multiple frequencies (annual and longer cycles) are triggered by astronomical solar-lunar quasi-periodic forcing, which are independent of the actual geographical arrangement of the continents and the oceans.

The existence of modern-like climatic oscillations like ENSO during the past can be well explained considering that they are largely regulated by external factors. The existence of important teleconnections between the main modern climate controls active in different hemispheres and oceans (Wyatt et al. 2011) further support our finding of an astronomical control of earth climate where configuration of the continents and the oceans likely does not produce the frequencies, but, at most, regulates the amplitudes of the corresponding cycles through internal resonances. Recently, Wang et al. (2012, their Figure 8) found that cycles at periods of 2.3, 3, 4.1, 5, 6.3, 7, 9-10, and 15-16 years, which are reasonably consistent with our findings, could be induced by solar-lunar tidal gravitational waves. This result reinforces our interpretation that cycles of this kind of are commonly found in both ancient and modern climate records because they are linked to almost stable astronomical cycles.

## 8. CONCLUSIONS

Our study of two evaporitic varved sequences in Sicily reveals that during the acme of the Messinian salinity crisis the deposition of the evaporites was controlled by quasi-periodic climate oscillations similar to those commonly described for modern analogs. The periodicities recognized in our study resemble the ones associated to the Quasi-Biennial Oscillation, the El Niño Southern Oscillation, the sunspot-number solar cycle and, interestingly, the 9 year periodicity related to a lunisolar tidal cycle. All of these represent the main climate forcing mechanisms that regulated temperature and precipitation in the Mediterranean in the past 500 years (Camuffo et al. 2010; Scafetta, 2010, 2012a; Wang et al, 2012). We suggest that during the MSC, at least during the insolation minima when evaporites precipitated, the Mediterranean did not have a constant hydrological budget but was rather characterized by a highly dynamic climate. In such a dynamic scenario the conditions of the Mediterranean seasonally switched between oversaturated evaporative and diluted non-evaporative conditions. This implies that, even during insolation minima, evaporitic conditions were achieved and lasted only for short time intervals (i.e., at seasonal scale) and not for the entire duration of the arid portion of the insolation cycle.

Moreover, we suggest that the role of astronomical high-frequency forcing, as from lunar and solar periodicities, can be reasonably extended back to the Late Miocene, or maybe even to more ancient periods, as demonstrated for the longer-term ones (precession, obliquity, eccentricity), because they are independent of local factors such as basin physiography and ocean and atmospheric circulation.

**ACKNOWLEDGMENTS**
This study received financial support by the Italian Minister of University and Research (MIUR) PRIN 2008 research project "High-resolution stratigraphy, palaeoceanography and palaeoclimatology of the Mediterranean area during the Messinian salinity crisis", coordinated by M. Roveri. Julien Gargani and an anonymous reviewer are greatly acknowledged for their helpful revisions. We would like to acknowledge also Associate Editor Kate Giles and Journal Editor Gene Rankey for their suggestions and comments which greatly improved this manuscript.

## REFERENCES CITED

ANDERSON, R.Y., 1982, Long geoclimatic record from the Permian: Journal of Geophysal Research, v. 87, p. 7285–7294.

ANDERSON, R.Y., AND DEAN, W.E., 1995, Filling the Delaware Basin: Hydrologic and climatic controls on the Upper Permian Castile Formation varved evaporite, *in* Scholle, P.A., Peryt, T.M., and Ulmer-Scholl, D.S., eds., The Permian of Northern Pangea, Vol. 2: Sedimentary Basins and Economic Resources: Berlin, Springer-Verlag, p. 61-78.

BALDWIN, M.P., GRAY, L.J., DUNKERTON, T.J., HAMILTON, K., HAYNES, P.H., RANDEL, W.J., HOLTON, J.R., ALEXANDER, M.J., HIROTA, I., HORINOUCHI, T., JONES, D.B.A., KINNERSLEY, J.S., MARQUARDT, C., SATO, K. AND TAKAHASHI, M., 2001, The Quasi-Biennial Oscillation: Reviews of Geophysics, v. 39, p. 179-229.

BLANC, P.L., 2000, Of sills and straits: a quantitative assessment of the Messinian Salinity Crisis: Deep-Sea Research, v. 47, Part I, p. 1429-1460.

BOURILLOT, R., VENIN, E., ROUCHY, J.-M., DURLET, C., ROMMEVAUX, V., KOLODKA., C., AND KNAP, F., 2009, Structure and evolution of a Messinian mixed carbonate-siliciclastic platform: the role of the evaporites (Sorbas Basin, South-east Spain): Sedimentology, v. 57- 2, p. 477–512.

CAMUFFO, D., BERTOLIN, C., DIODATO, N., BARRIENDOS, M., DOMINGUEZ-CASTRO, F., COCHEO, C. DELLA VALLE, A., GARNIER, E. AND ALCOFORADO, M.-J., 2010, The western Mediterranean climate: how will it respond to global warming?: Climatic Change, v. 100, p. 137-142.

CIESM, 2008, The Messinian Salinity Crisis mega-deposits to microbiology - A consensus report: Monaco, CIESM Workshop Monograph, v. 33, p. 73–82.

COURTILLOT, V., LE MOUËL, J. L., AND MAYAUD, P.N., 1977, Maximum entropy spectral analysis of the geomagnetic activity index aa over a 107-Year Interval: Journal of Geophysical Research, v. 82-19, p. 2641-2649.

DAVIES, A., KEMP, A.E.S., WEEDON, G.P., AND BARRON, J.A., 2012, El Niño Southern Oscillation variability from the Late Cretaceous Marca Shale of California: Geology, v. 40-1, p.15-18.

DECIMA, A., AND WEZEL, F.C., 1971, Osservazioni sulle evaporiti Messiniane della Sicilia centro-meridionale: Rivista Mineraria Siciliana, v. 130–134, p. 172–187.

DUGGEN, S., HOERNLE, K., VAN DEN BOGAARD, P., RÜPKE,L. AND PHIPPS MORGAN, J., 2003. Deep roots of the Messinian salinity crisis, Nature, v. 422, p. 602-606.

EHRET, T., 2008, Old brass brains-mechanical prediction of tides: American Congress on Surveying and Mapping, Bulletin, v. 6, p. 41–44.

FEDOROV, A.V., DEKENS, P.S., MCCARTHY, M., RAVELO, A.C., DEMENOCAL, P.B., BARREIRO, M., PACANOWSKI, R.C. AND PHILANDER, S.G., 2006, The Pliocene paradox (mechanisms for a permanent El Niño): Science, v. 312, p. 1485-1489.

FLECKER, R., DE VILLIERS, S., AND ELLAM, R.M., 2002, Modelling the effect of evaporation on the salinity $^{87}Sr/^{86}Sr$ relationship in modern and ancient marginal-marine systems: the Mediterranean Messinian Salinity Crisis: Earth and Planetary Science Letters, v. 203, p. 221–233.

FLECKER, R., AND ELLAM, R.M., 2006, Identifying Late Miocene episodes of connection and isolation in the Mediterranean-Paratethyan realm using Sr isotopes: Sedimentary Geology, v. 188–189, p. 189–203.

GALEOTTI, S., VON DER HEYDT, A., HUBER, M., BICE, D., DIJKSTRA, H., JILBERT, T., LANCI, L., AND

Reichart, G-J, 2010, Evidence for active El Niño Southern Oscillation variability in the Late Miocene greenhouse climate: Geology, v. 38-5, p. 419–422.

Gargani, J., Moretti, I. , and Letouzey, J., 2008, Evaporite accumulation during the Messinian Salinity Crisis: The Suez Rift case: Geophysical Research Letters, v. 35, p. L02401, doi:10.1029/2007GL032494.

Gargani, J.,and Rigollet, C., 2008, Mediterranean sea level variations during the Messinian salinity crisis: Geophysical Research Letters, v. 34, p. L10405, doi:10.1029/2007GL029885.

Govers, R., Meijer P., and Krijgsman W., 2008, Regional isostatic response to Messinian Salinity Crisis events: Tectonophysics, v. 463, p. 109-129.

Hoyt, D.V., and Schatten K.H., 1997, The Role of the Sun in the Climate Change: New York, Oxford Univ. Press, 279 p.

Hsü, K.J., Montadert, L., Bernoulli, D., Cita, M.B., Erickson, A., Garrison, R.E., Kidd, R.B., Méliéres, F., Mœller, C. and Wright, R. 1978, History of the Mediterranean salinity crisis: Nature, v. 267, p. 399-403.

Hsü, K.J., Ryan, W.B.F., and Cita, M.B., 1973, Late Miocene desiccation of the Mediterranean: Nature, v. 242, p. 240-244.

Hutton, J., 1785, Abstract of a dissertation read in the Royal Society of Edinburgh, upon the seventh of March, and fourth of April, MDCCLXXXV, Concerning the System of the Earth, Its Duration, and Stability. Edinburgh. 30 p.

Hutton, J., 1795, Theory of the Earth; With Proofs and Illustrations: Edinburgh, Creech, 2 vols, 126 p.

Jolivet, L., Augier, R., Robin, C., and Suc, J.P., 2006, Lithospheric-scale geodynamic context of the Messinian salinity crisis: Sedimentary Geology, v. 188-189, p. 9-33.

Kirkland, D.W., 2003, An explanation for the varves of the Castile evaporites (Upper Permian), Texas and New Mexico, USA: Sedimentology, v. 50, p. 899-920.

Kirkland, D.W., Denison, R.E., and Dean, W.E., 2000, Parent brine of the Castile evaporites (Upper Permian), Texas and New Mexico: Journal of Sedimentary Research, v. 70, p. 749–761.

Krijgsman, W., Hilgen, F.J., Raffi , I., Sierro, F.J., and Wilson, D.S., 1999, Chronology, causes, and progression of the Messinian salinity crisis: Nature, v. 400, p. 652–655.

Leslie, A.B., Kendall, A.C., Harwood, G.M. and Powers, D.W., 1996, Conflicting indicators of palaeodepth during deposition of the Upper Permian Castile Formation, Texas and New Mexico: *in* Kemp, A.E.S., Ed., Palaeoclimatology and Palaeoceanography from Laminated Sediments, Geological Society of London, special publication 116, p. 79-92.

Lugli, S., 1999, Geology of the Realmonte salt deposit, a desiccated Messinian Basin (Agrigento, Sicily): Società Geologica Italiana, Memorie, v. 54, p. 75-81.

Lugli S., Schreiber B.C. and Triberti, B., 1999, Giant polygons in the Realmonte mine (Agrigento, Sicily): evidence for the desiccation of a Messinian halite basin: Journal of Sedimentary Research, v. 69, p. 764-771.

Lugli, S., Bassetti, M.A., Manzi, V., Barbieri, M., Longinelli, A., Roveri, M. and Ricci Lucchi, F., 2007, The Messinian “Vena del Gesso” evaporites revisited: isotopic and organic matter characterization: *in* Schreiber, B.C., Lugli, S. and Babel, M.: Evaporites Through Space and Time, Geological Society of London, Journal, v. 285, p. 143–154.

Lugli, S., Manzi, V., Roveri, M., and Schreiber, B.C., 2010, The Primary Lower Gypsum in the Mediterranean: A new facies interpretation for the first stage of the Messinian salinity crisis: Palaeogeography, Palaeoclimatology, Palaeoecology, v. 297, p. 83–99.

Lyell, C., 1830, Principles of geology, being an attempt to explain the former changes of the Earth's surface, by reference to causes now in operation: London, John Murray, v. 1, p. 512.

Manzi, V., Roveri, M., Gennari, R., Bertini, A., Biffi , U., Giunta, S., Iaccarino, S.M., Lanci, L., Lugli, S., Negri, A., Riva, A., Rossi, M.E., and Taviani, M., 2007, The deep-water counterpart of the Messinian Lower Evaporites in the Apennine foredeep: The Fanantello section (northern Apennines, Italy): Palaeogeography, Palaeoclimatology, Palaeoecology, v. 251, p. 470–499, doi: 10.1016/j.palaeo.2007.04.012.

Manzi, V., Lugli, S., Roveri, M., and Schreiber, B.C., 2009, A new facies model for the Upper Gypsum of Sicily (Italy): Chronological and paleoenvironmental constraints for the Messinian salinity crisis in the Mediterranean: Sedimentology, v. 56, p. 1937–1960.

MANZI, V., GENNARI, R., LUGLI, S., ROVERI, M., AND SCHREIBER, B.C., 2011, The Messinian "Calcare di Base" (Sicily, Italy) revisited. Geological Society of America, Bulletin, v. 123-1/2, v. 347–370.

MEIJER, P., AND KRIJGSMAN, W., 2005, A quantitative analysis of the desiccation and re-filling of the Mediterranean during the Messinian Salinity Crisis: Earth and Planetary Science Letters, v. 240, p. 510–520

MAZZARELLA, A., AND N., SCAFETTA, 2012, Evidences for a quasi 60-year North Atlantic Oscillation since 1700 and its meaning for global climate change: Theoretical and Applied Climatology, v. 107, p. 599-609.

MONTADERT, L., J. LETOUZEY, AND A. MAUFFRET, 1978, Messinian event: Seismic evidence, Initial Reports Deep See Drilling Project, v. 42-1, p. 1037–1050, doi:10.2973/dsdp.proc.42-1.154.

MUÑOZ, A., OJEDA, J., AND SÁNCHEZ-VALVERDE, B., 2002, Sunspot-like and ENSO/NAO-like periodicities in lacustrine laminated sediments of the Pliocene Villarroya Basin (La Rioja, Spain): Journal of Paleolimnology, v. 27, p. 453–463.

OGNIBEN, L., 1957, Petrografia della serie solfifera-siciliana e considerzioni geotecniche relative: Memorie Descrittive della Carta Geologica d'Italia, v. 33, 275 p.

OGURTSOV, M.G., NAGOVITSYN, Y.A., KOCHAROV, G.E., AND JUNGNER, H., 2002, Long-period cycles of the Sun's activity recorded in direct solar data and proxies: Solar Physics, v. 211, p. 371-394.

PRESS, W.H., TEUKOLSKY, S.A., VETTERLING, W.T. AND FLANNERY, B.P., 1997, Numerical recipes in C: The art of Scientific Computing, Second edition: Cambridge U.K., Cambridge University Press, 994 p.

RAYMO, M.E., RUDDIMAN, W.F., BACKMAN, J., CLEMENT, B.M., AND MARTINSON, D.G., 1989, Late Pliocene variation in northern hemisphere ice sheets and North Atlantic deep water circulation: Paleoceanography, v. 4, p. 413–446.

ROUCHY, J.M., AND CARUSO, A., 2006, The Messinian salinity crisis in the Mediterranean basin: a reassessment of the data and an integrated scenario: Sedimentary Geology, v. 188–189, p. 35–67.

ROVERI, M., AND TAVIANI, M., 2003, Calcarenite and sapropel deposition in the Mediterranean Pliocene: shallow- and deep-water record of astronomically driven climatic events: Terra Nova, v. 15, p. 279-286.

ROVERI, M., LUGLI, S., MANZI, V., AND SCHREIBER, B.C., 2008, The Messinian Sicilian stratigraphy revisited: Toward a new scenario for the Messinian salinity crisis: Terra Nova, v. 20, p. 483–488.

ROVERI, M., GENNARI, R., LUGLI, S., AND MANZI, V., 2009, The Terminal Carbonate Complex: the record of sea-level changes during the Messinian salinity crisis: GeoActa, v. 8, p. 57-71

SCAFETTA, N., 2010, Empirical evidence for a celestial origin of the climate oscillations and its implications: Journal of Atmospheric and Solar-Terrestrial Physics, v. 72, p. 951–970.

SCAFETTA N., 2012a, A shared frequency set between the historical mid-latitude aurora records and the global surface temperature: Journal of Atmospheric and Solar-Terrestrial Physics, v. 74, p. 145-163.

SCAFETTA N., 2012b, Testing an astronomically based decadal-scale empirical harmonic climate model versus the IPCC (2007) general circulation climate models: Journal of Atmospheric and Solar-Terrestrial Physics, v. 80, p. 124-137.

SCAFETTA N., 2012c, Multi-scale harmonic model for solar and climate cyclical variation throughout the Holocene based on Jupiter-Saturn tidal frequencies plus the 11-year solar dynamo cycle: Journal of Atmospheric and Solar-Terrestrial Physics, v. 80, p. 296-311.

SCAFETTA N., 2012d, Does the Sun work as a nuclear fusion amplifier of planetary tidal forcing? A proposal for a physical mechanism based on the mass-luminosity relation: Journal of Atmospheric and Solar-Terrestrial Physics, v. 81-82, p. 27-40.

SCHREIBER, B.C., HSÜ, K.J., 1980. Evaporites: *in* Hobson, G.D. ed., Applied Science Publishers, London, Developments in Petroleum Geology, v. 2., p. 87–138.

SELLI, R., 1960, Il Messiniano Mayer-Eymar 1867: Proposta di un neostratotipo: Giornale di Geologia, v. 28, p. 1–33.

TRENBERTH, K.E., 1997, The definition of El Niño: American Meteorological Society, Bulletin, v. 78, p. 2771-2777.

VAN COUVERING, J.A., CASTRADORI, D., CITA, M.B., HILGEN, F.J., AND RIO, D., 2000, The base of the Zanclean Stage and of the Pliocene Series: Episodes, v. 23-3, p. 179–187.

VAN DER LAAN, E., SNEL, E., DE KAENEL, E., HILGEN, F.J., AND KRIJGSMAN, W., 2006, No major deglaciation across the Miocene-Pliocene boundary: Integrated stratigraphy and astronomical tuning of the Loulja sections (Bou Regreg area, NW Morocco): Paleoceanography, v. 21, p. PA3011, doi:10.1029/2005PA001193.

WANG, Z., WU, D., SONG, X., CHEN, X., AND NICHOLLS, S., 2012, Sun-Moon gravitation-induced wave characteristics and climate variation: Journal of Geophysical Research, v. 117, p. D07102, doi:10.1029/2011JD016967.

WARA, M.W., RAVELO, A.C., AND DELANEY, M.L., 2005, Permanent El Niño-like conditions during the Pliocene warm period: Science, v. 309, p. 758–761.

WARREN, J.K., 2006, Evaporites: Sediments, Resources and Hydrocarbons: Berlin, Springer, 1036 p.

WATANABE, T., SUZUKI, A., MINOBE, S., KAWASHIMA, T., KAMEO, K., MINOSHIMA, K., AGUILAR, Y. M., WANI, R., KAWAHATA, H., SOWA, K., NAGAI, T., AND KASE, T., 2011, Permanent El Niño during the Pliocene warm period not supported by coral evidence: Nature, v. 471, p. 209-211.

WYATT, M.G., KRAVTSOV, S., AND TSONIS, A.A., 2011, Atlantic Multidecadal Oscillation and Northern Hemisphere climate variability: Climate Dynamics, v. 38, p. 929-949.

YNDESTAD, H., 2006, The influence of the lunar nodal cycle on Arctic climate: Journal of Marine Science, v. 63, p. 401-420.

TABLE 1.—*Major climate cycles and their possible astronomical origin.*

| Period [yr] | | Cause of Climate Variability | Main Effects on Earth Climate | References |
|---|---|---|---|---|
| 2.3–2.5 | QBO | quasi-biennial oscillation | Monsoon intensity | Baldwin 2001 |
| 3–7 | ENSO | El Niño Southern Oscillation | Pacific Ocean temperature<br>Pacific Ocean circulation<br>Global surface temperature,<br>Global rainfall | Trenberth 1997 |
| ≈ 9 | SLC | Lunisolar long-term tidal cycle | Global surface temperature<br>Global rainfall | Scafetta 2010, 2011 |
| 10–13 | 11-SSC | 11-year sunspot number solar cycle | Global surface temperature,<br>Ocean temperature | Hoyt and Schatten 1997 |
| 18–25 | 22-HSC<br>20-JSHC<br>18.6-SLC | 22 yr Hale solar magnetic cycle<br>20 yr Jupiter/Saturn heliospheric cycle<br>18.03 yr Saros cycle<br>18.6 yr lunisolar nodal cycle<br>18.6 yr nutation cycle | Monsoon intensity<br>Global surface temperature<br>Influence on Arctic climate<br>Regulation of bidecadal North Atlantic Oscillation (NAO) | Hoyt and Schatten 1997<br>Scafetta 2010, 2011<br>Yndestad 2006 |
| 60–100 | 80-GSC<br>60-JSHC | 80-Gleissberg solar cycle<br>60-yr Jupiter/Saturn heliospheric cycle | Global surface temperature<br>Global rainfall | Hoyt and Schatten 1997<br>Scafetta 2010, 2011 |

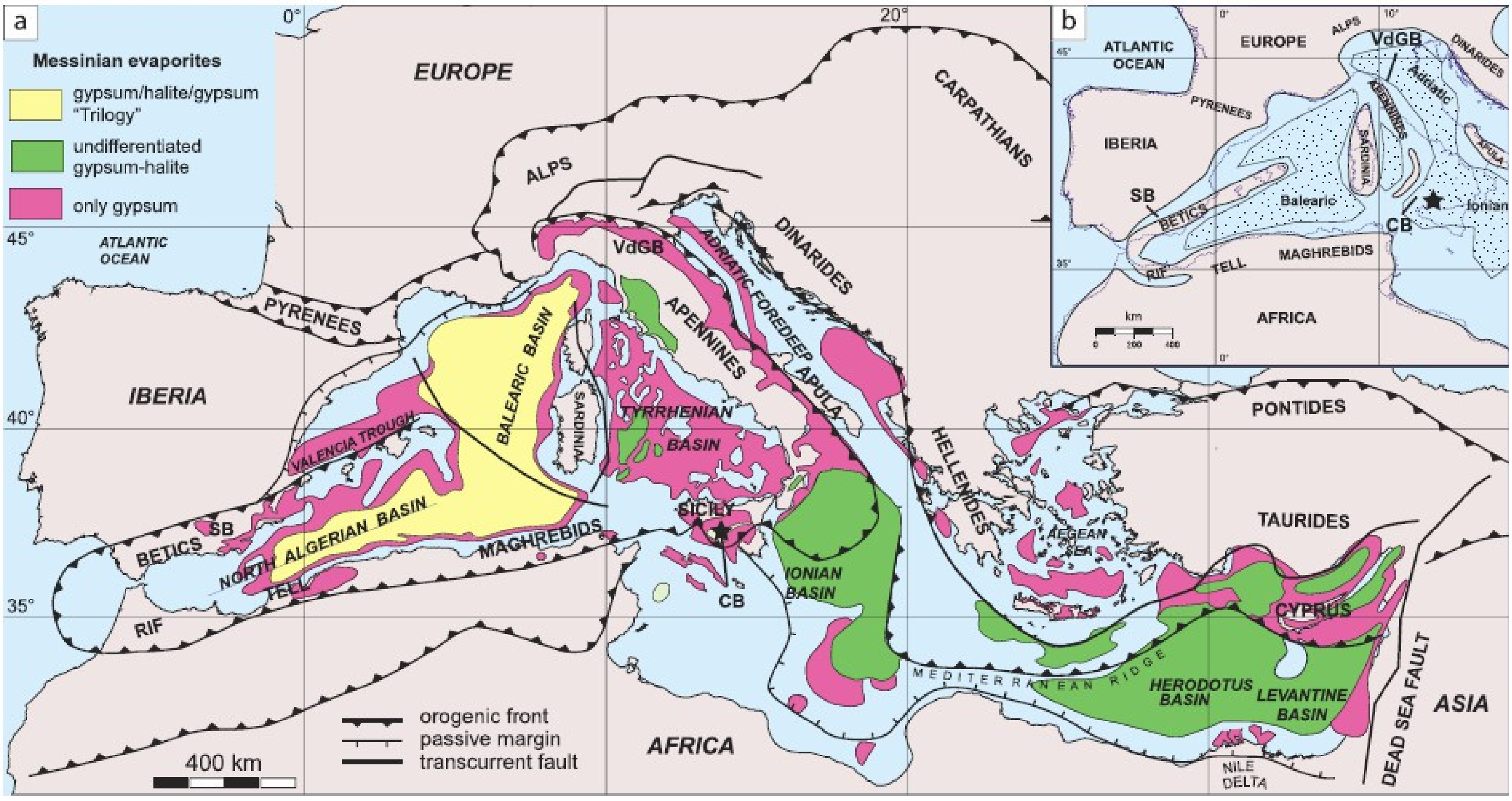

FIG.1) A) Map of the Messinian evaporites in the Mediterranean modified after Rouchy and Caruso (2006). The term "trilogy" indicates the threefold deeper succession of Western Mediterranean that, based on seismic, includes a halite unit sandwiched between two gypsum units. Alternatively when both halite and gypsum are present but not distinguishable, the term "undifferentiated" has been adopted. B) Paleocenaographic map of the Western Mediterranean basins during the Messinian salinity crisis, showing the main evaporite depocentres (dotted areas). Modified after Jolivet et al. (2006). Emerged areas are in gray; dotted line is the modern coastline; star indicates the study area; CB, Caltanissetta basin; SB, Sorbas basin; VdGB, Vena del Gesso basin.

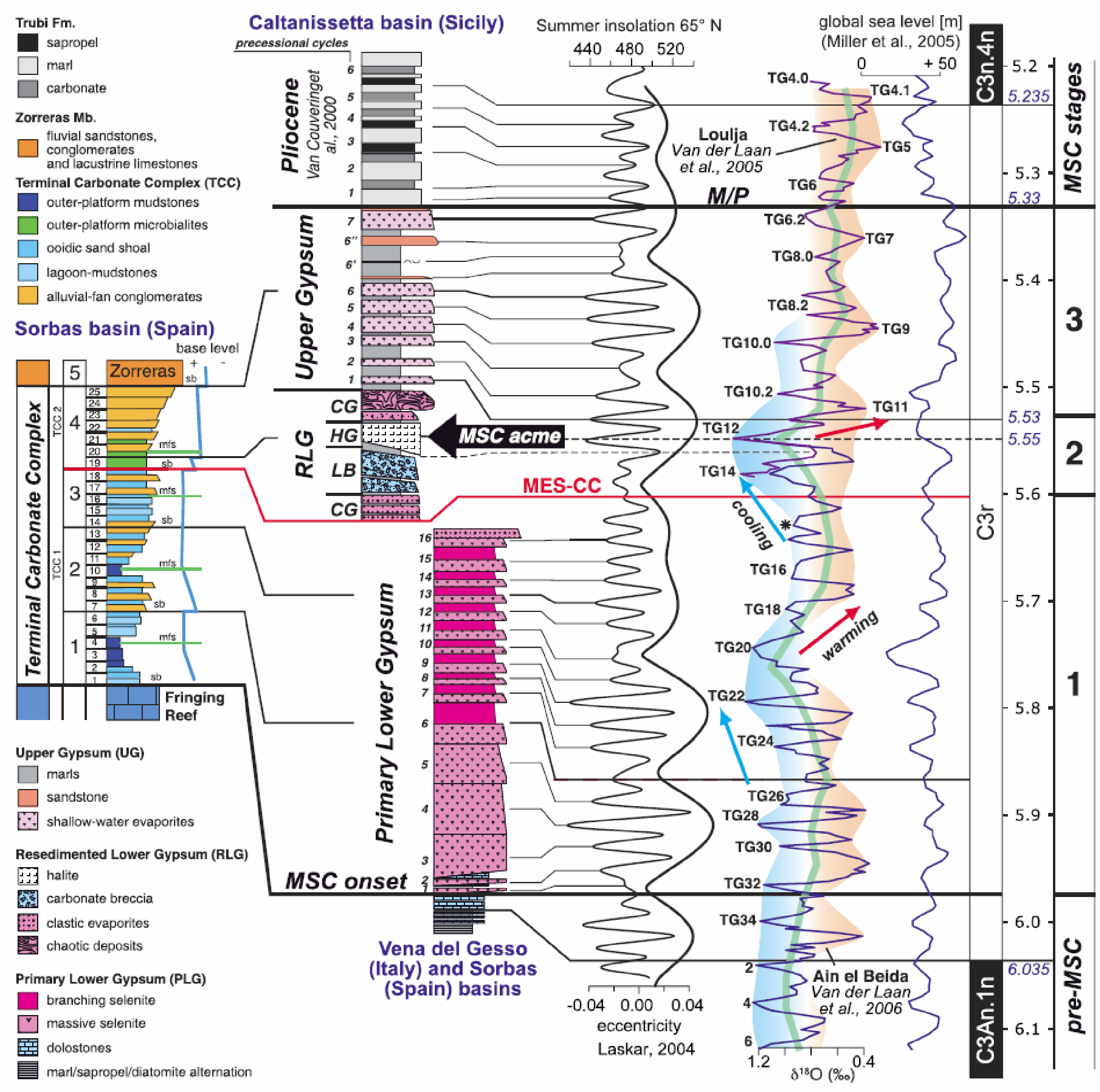

FIG.2) Stratigraphic framework of the Messinian salinity crisis with the indication of the various evaporite deposits and the key surfaces. The Primary Lower Gypsum unit (PLG, Lugli et al. 2010) was deposited during stage 1 of MSC. Deposited during Stage 2, the Resedimented Lower Gypsum unit (RLG) is a complex unit including clastic gypsum (CG), halite, and gypsum primary cumulate deposits (HG); and clastic carbonates derived from evaporitic and microbial limestone (LB, limestone breccia; Calcare di base type 3, *sensu* Manzi et al. 2011). Finally, during stage 3 a further primary gypsum unit was deposited (UG, Upper Gypsum; see Manzi et al. 2009). The carbonate unit of the Terminal Carbonate Complex (TCC; Bourillot et al. 2009; Roveri et al. 2009) are also included: **mfs** (maximum flooding surface); **sb** (sequence boundary); 1-4 (transgressive-regressive large-scale sequences, ~ 100 kyr). MSC key-surfaces are: **MSC onset** (start of evaporite deposition in the Mediterranean); **MES-CC** (correlative conformity of the Messinian erosional surface; see Roveri et al. 2008); **M/P** (Miocene-Pliocene boundary).

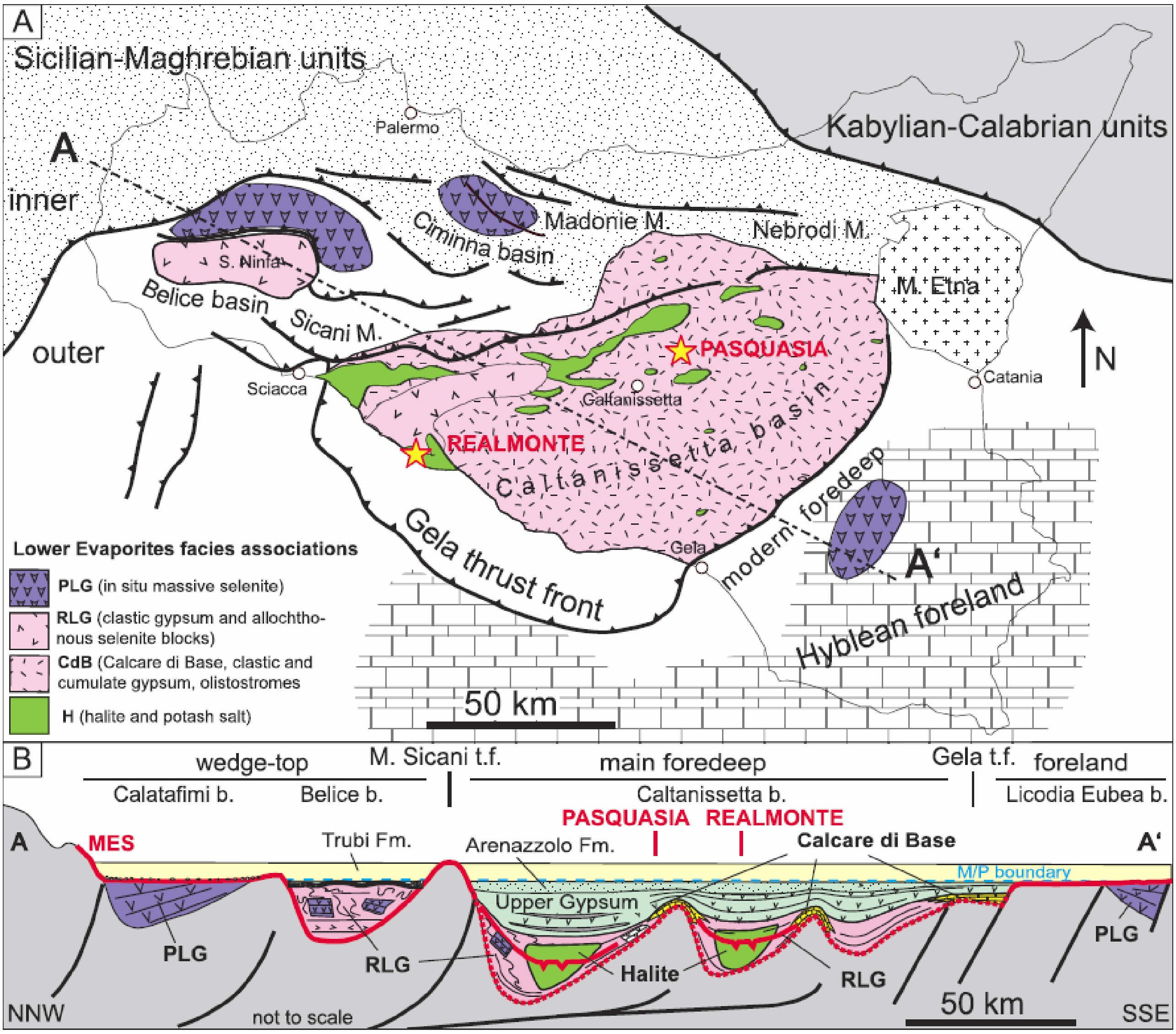

FIG.3) A)Schematic geological map of Sicily (modified after Manzi et al. 2011) with the location of the Pasquasia and Realmonte sections. B) Schematic geological cross section across the Sicilian Basin (modified from Roveri et al. 2008). **PLG**, Primary Lower Gypsum; **RLG**, Resedimented Lower Gypsum; **H**, Halite unit; **UG**, Upper Gypsum.

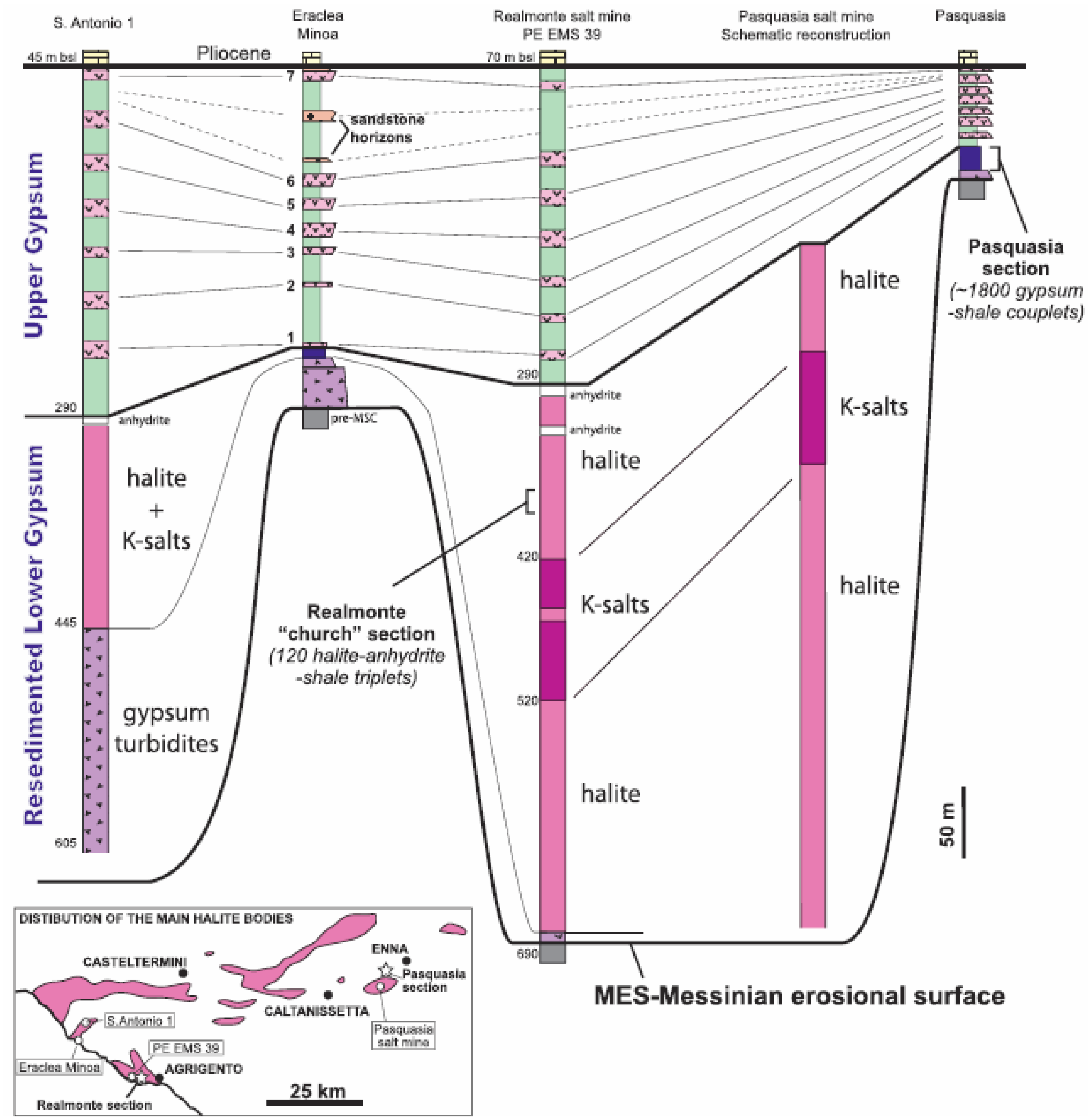

FIG.4) Stratigraphic correlation in the Caltanissetta basin. Note the quasi-symmetrical pattern of the halite unit. Eraclea Minoa and Pasquasia sections are from Manzi et al. (2009). S.Antonio 1 and PE EMS 39 boreholes unpublished data, courtesy of ITALKALI SpA.

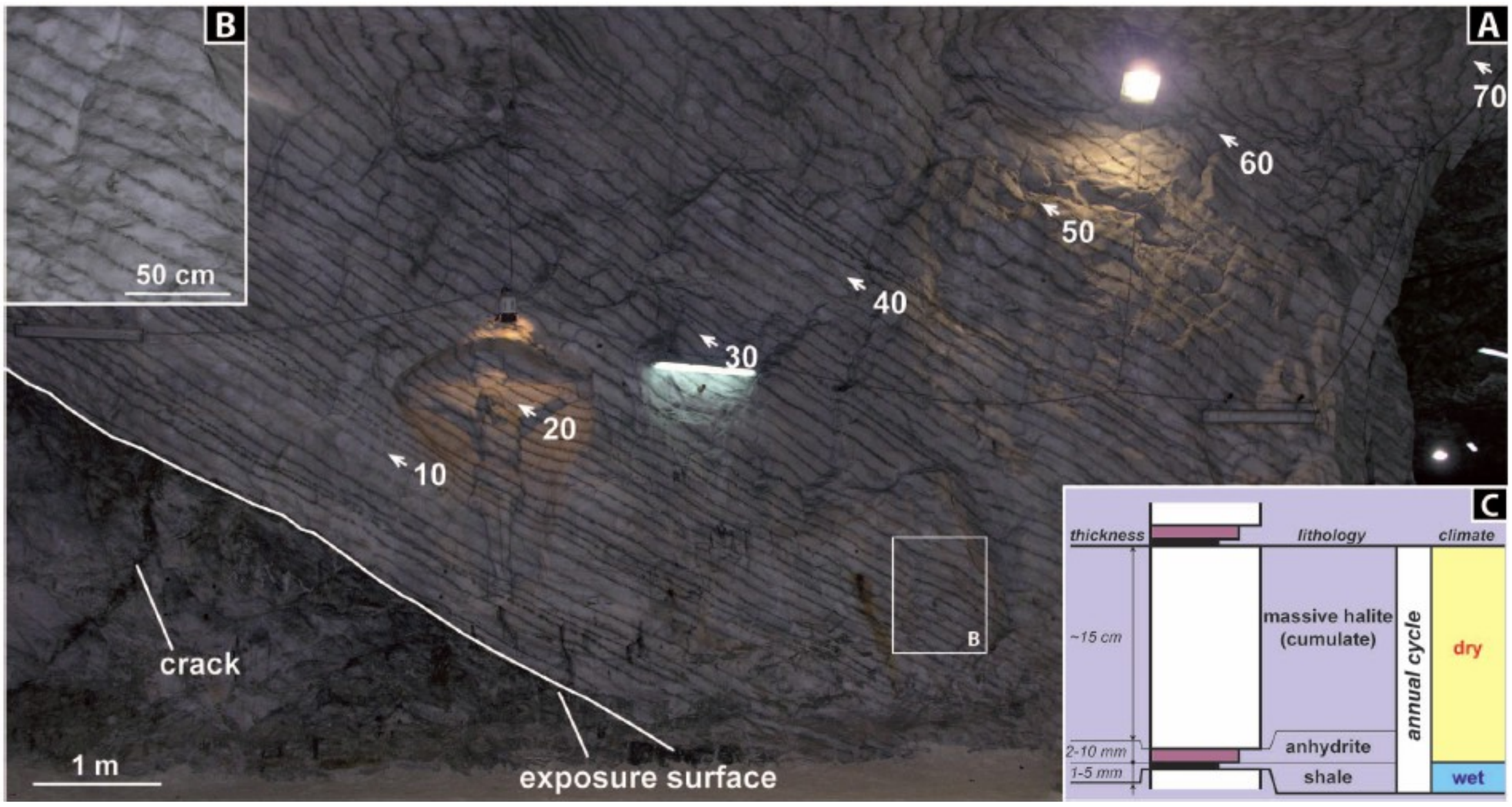

FIG.5) The “Church” section at the Realmonte Mine, Caltanissetta basin, Sicily. Small-scale cyclicity in the halite body included in the Resedimented Lower Gypsum unit. The lowermost 70 cycles of the section. A),number of halite varves in white. B) Closer view enlargement of inset in part A and C) schematic description of the tripartite cycles formed from the base by: i) millimeter-thin shale veneer; ii) millimeter- to centimeter-thick anhydrite and/or polyhalite layer; iii) decimeter-thick massive halite (white layers). See text for further information.

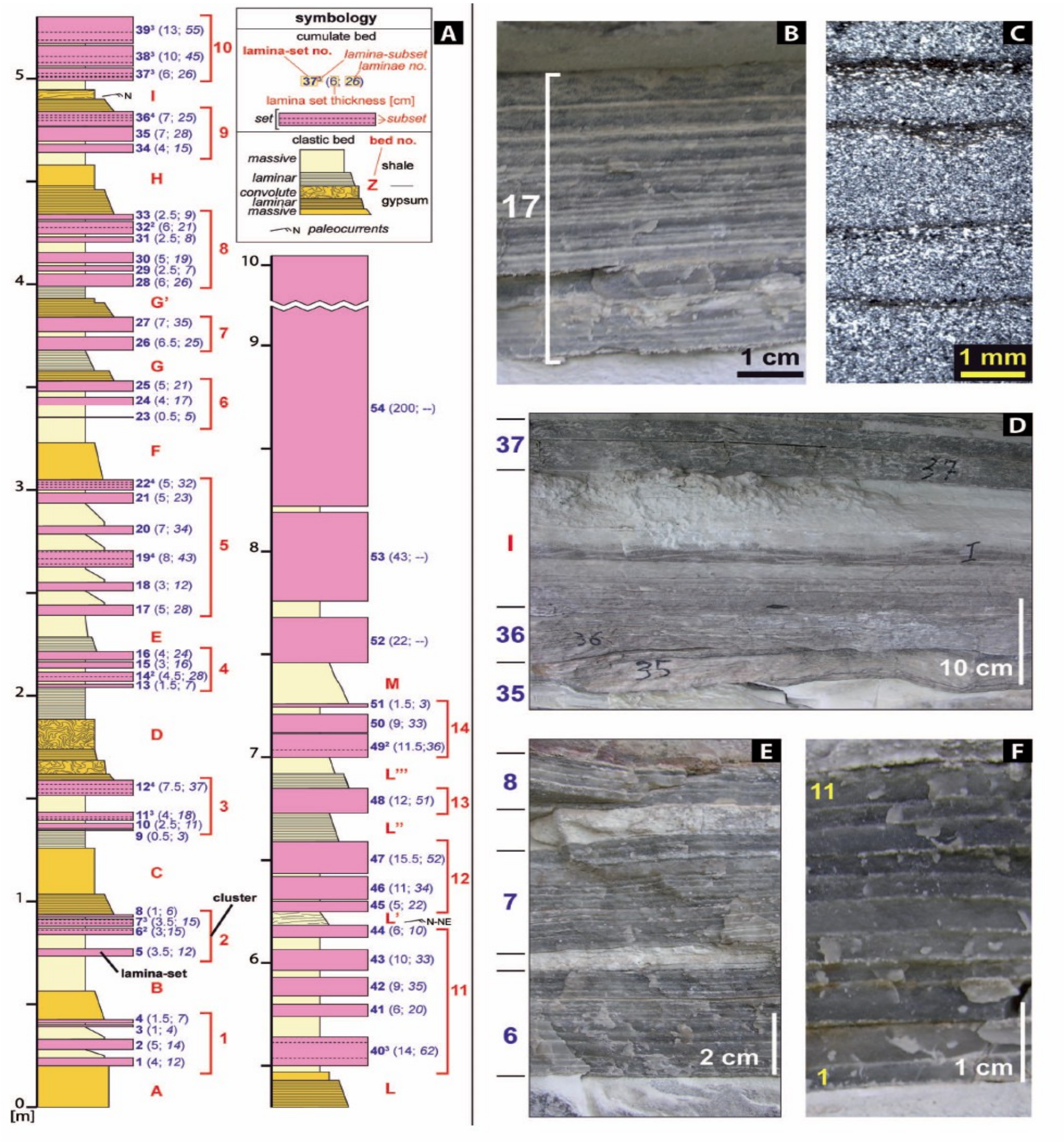

FIG.6) Small-scale cyclicity in the gypsum cumulates included in the Resedimented Lower Gypsum unit, Pasquasia, Caltanissetta Basin, Sicily. A) sedimentological log showing cumulate gypsum lamina sets (1 to 54) grouped to form clusters separated by low-density gypsum turbidite layers (A to M); internal subdivisions used in the definition of the 79 lamina subsets are indicated by dashed lines. B) Close view of cluster 17. Note the almost regular variation of the thickness cumulate laminae. C) Photomicrograph at crossed polars of the cumulate laminae in the lower part of cluster 17; note the distinctive inverse gradation. Each single lamina records a discrete (seasonal) evaporitc event as evidenced by the lack of internal subdivisions or inversions of crystal-size trend. D) Close view of the low-density turbidite "I" interlayered within the cumulate succession. No evidence of erosion has been observed at the base of the clastic bed. E) thin shale layers separating the lamina set; F) close view of lamina set 11.

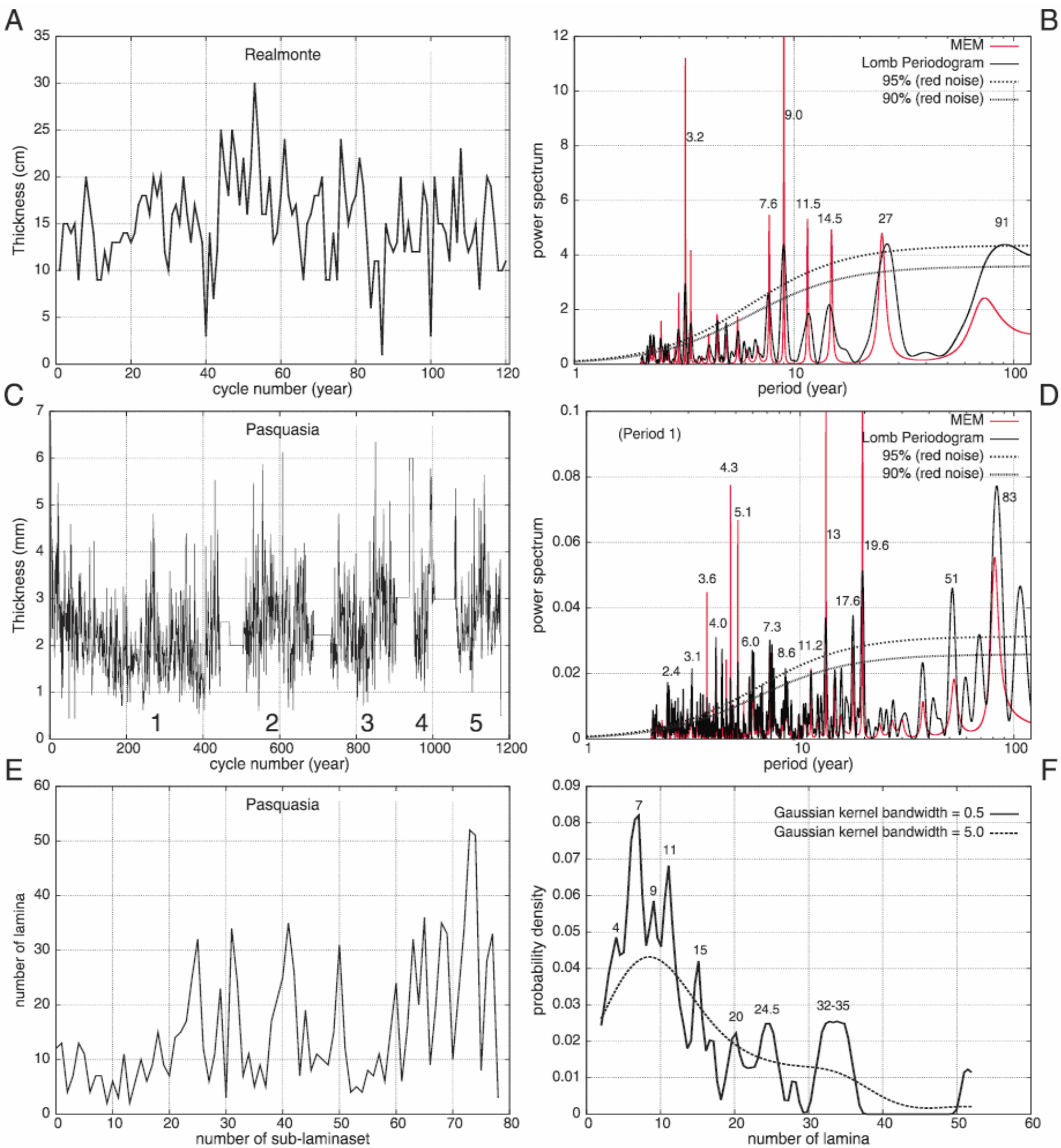

FIG.7) Statistical analyses performed in the primary deposit of the Resedimented gypsum unit. A) Thickness record of the salt layers in the "Church" section at the Realmonte salt mine. B) Maximum-entropy-method (MEM) analysis and Lomb periodogram of the Realmonte record. C) Thickness record of gypsum cumulate at Pasquasia. The record appears to be segmented into five groups because of regions that could not be annually resolved due to some measure of alteration; these intervals are filled with their average value. D) Maximum-entropy-method analysis and Lomb periodogram of the gypsum cumulates at Pasquasia. E) Number of laminae (and thus the number of years) forming each lamina set of the Pasquasia section. F) Recurrence periodicities of the dilution events in the basin obtained from probability distribution analysis of the duration of the 54 lamina sets and 79 lamina subsets of the Pasquasia section. Note that the shortest period of two years refers to the Nyquist frequency for annually resolved records, and that the spectral uncertainty increases with the period, as expected.

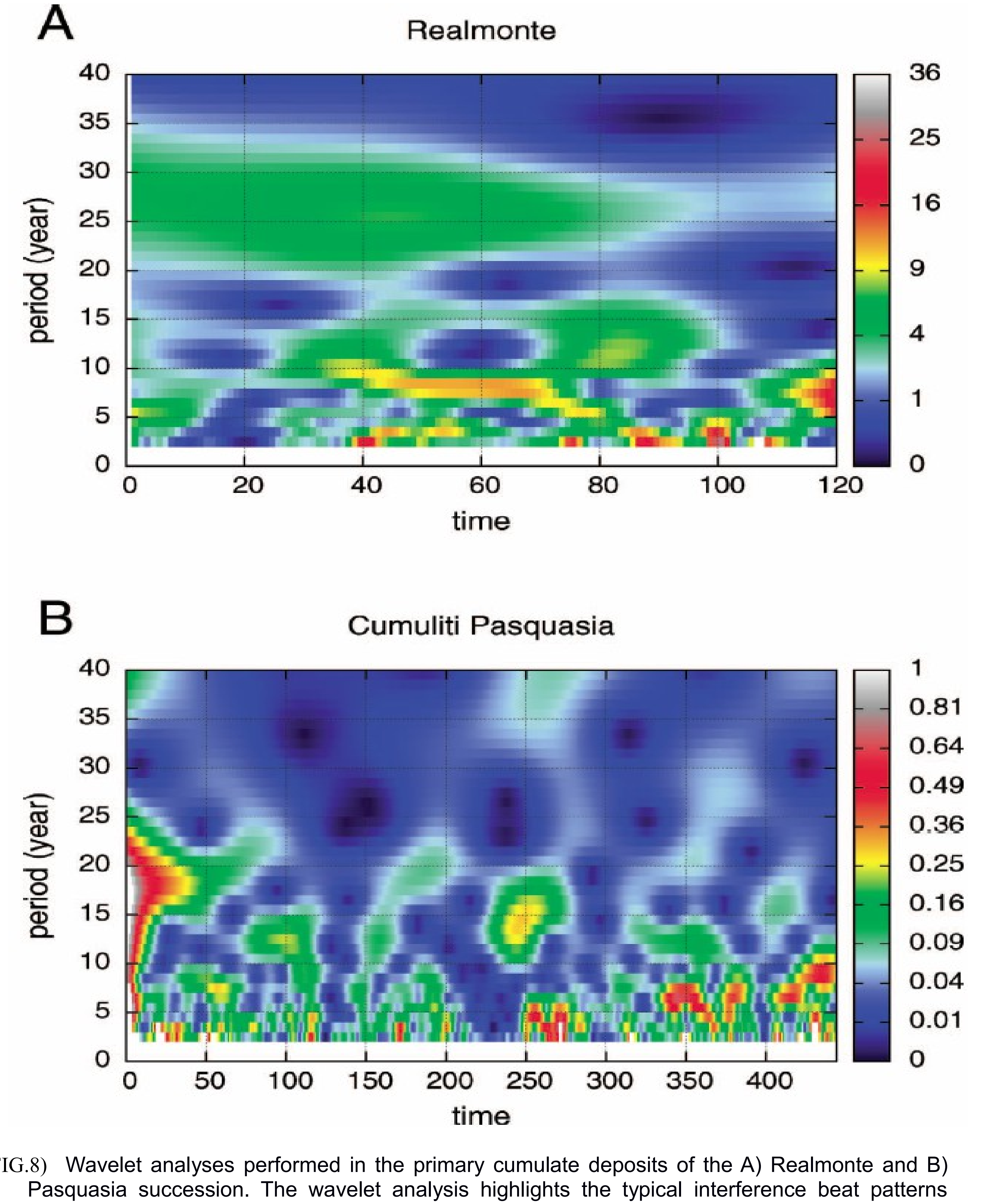

FIG.8) Wavelet analyses performed in the primary cumulate deposits of the A) Realmonte and B) Pasquasia succession. The wavelet analysis highlights the typical interference beat patterns among the frequencies that make up the records, and a stationary signal occuring between ~ 2 and ~ 7 years, possibly corresponding to El Niño, is visible. Given the short length of the data, results referring to periods larger than 20 years becomes progressively more uncertain.

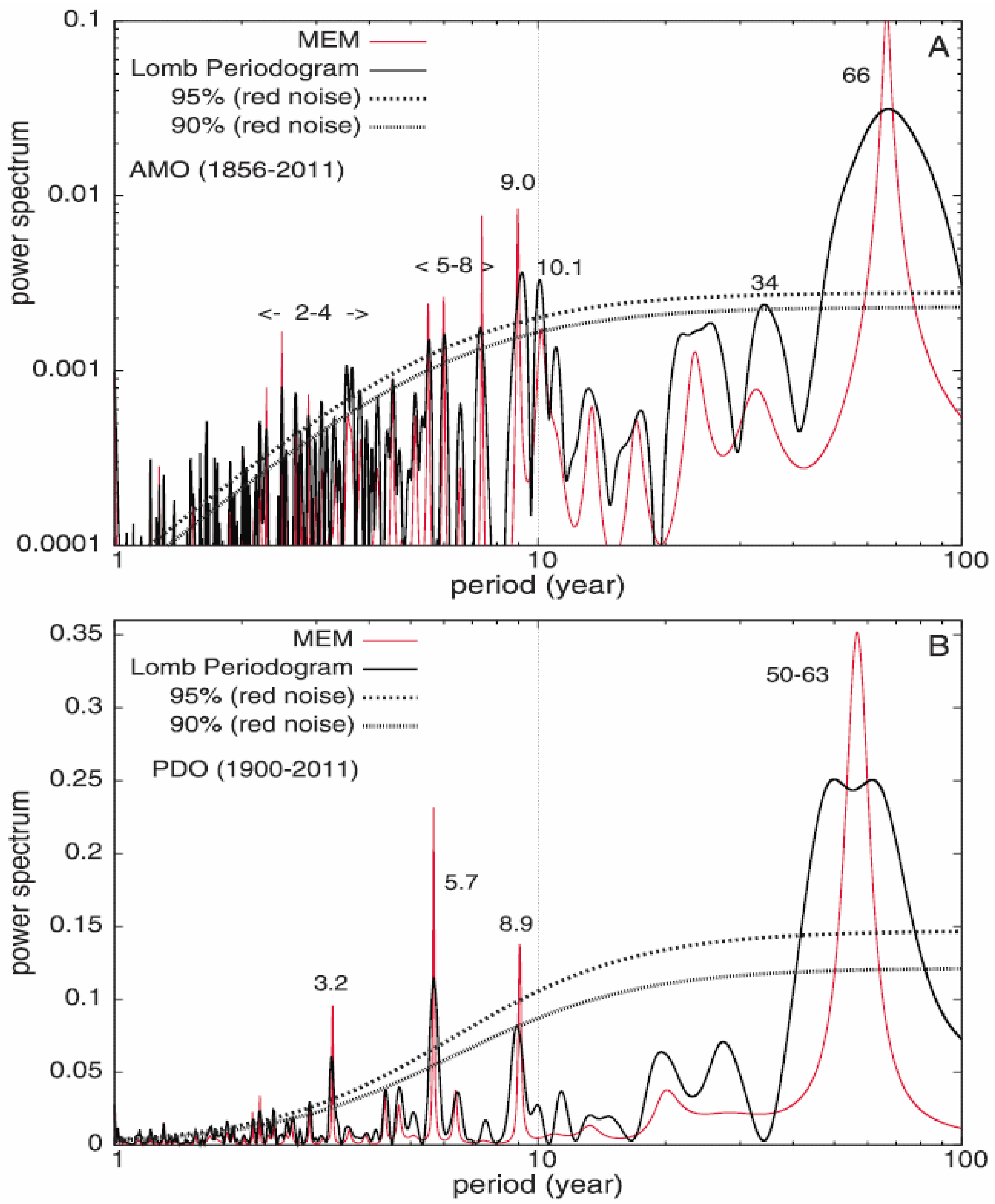

FIG.9) Maximum-entropy-power spectrum and Lomb periodogram estimates of (A) the Atlantic Multidecadal Oscillation (AMO; 1856-2011), and B) the pacific Decadal Oscillations (PDO; 1900-2011). Note that major peaks occur in the bands 2-3 years, 5-8 years, and about 9 years and about 60 years. (Data available via the internet at url: http://www.esrl.noaa.gov/psd/data/timeseries/AMO/ and http://www.atmos.washington.edu/~mantua/abst.PDO.html)

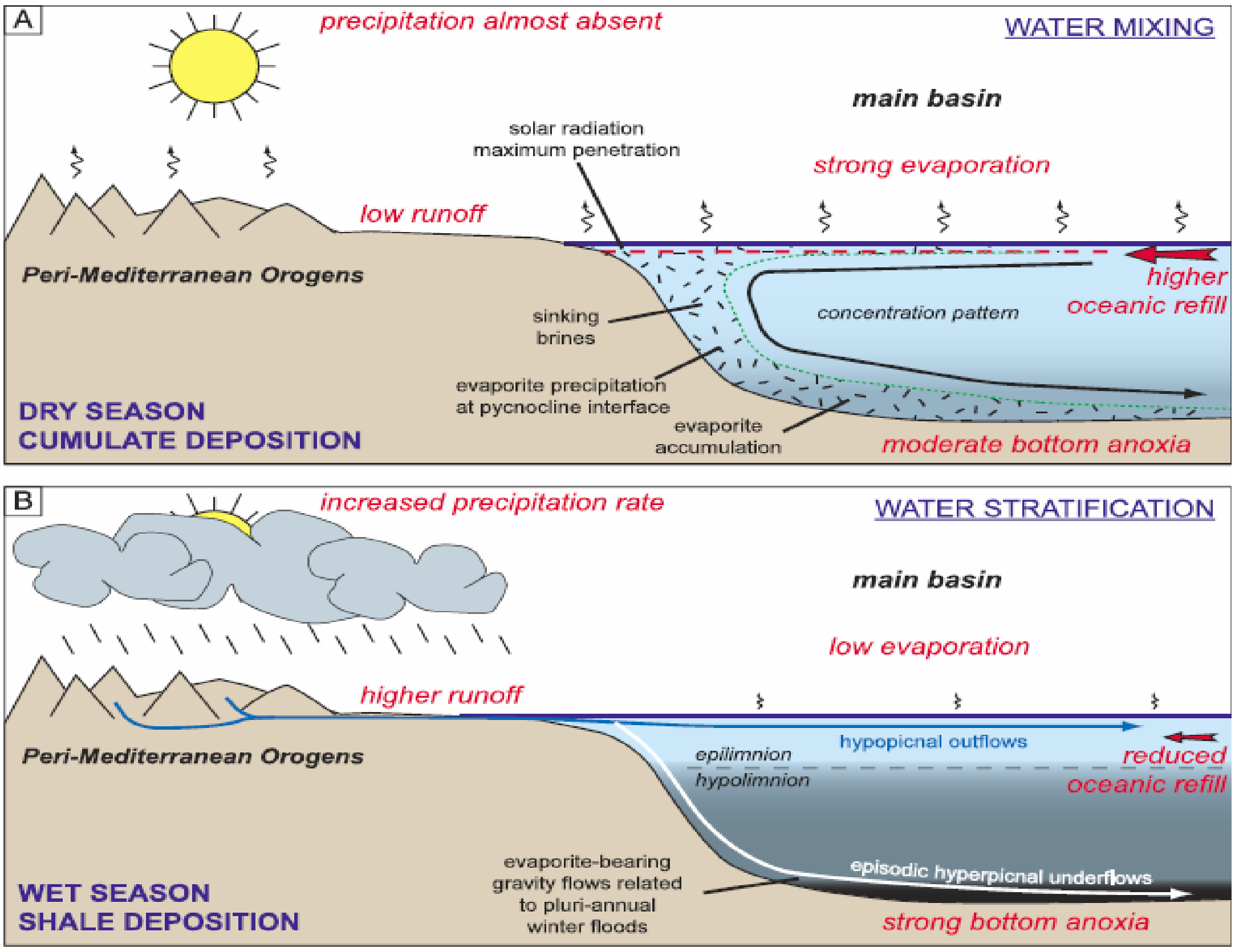

FIG.10) Schematic depositional model for the evaporite-shale annual couplets. A) Gypsum or anhydrite plus halite cumulates were deposited during dry seasons, B) shale during the wet season.